\documentclass[
journal=jacsat,
manuscript=communication,
amsmath,amssymb,
]{achemso}

\usepackage{chemformula} 
\usepackage[T1]{fontenc} 

\usepackage{natbib}

\usepackage{graphicx}
\graphicspath{{./}}
\usepackage{epstopdf}
\usepackage{psfrag}
\usepackage{dcolumn}

\usepackage{bm}
\usepackage{color}
\definecolor{darkblue}{rgb}{0,0,.8}
\usepackage{hyperref}
\hypersetup{pdftex=true, colorlinks=true, breaklinks=true, linkcolor=darkblue, citecolor=darkblue, menucolor=darkblue, pagecolor=darkblue, urlcolor=blue}

\usepackage[inline]{enumitem}
\usepackage{gensymb}

\usepackage[english]{babel}
\usepackage{listings}
\author{Florian Dei\ss enbeck}
\affiliation[MPIE]
{Max-Planck-Institut f\"ur Eisenforschung GmbH, Max-Planck-Stra{\ss}e 1, 40237 D\"usseldorf, Germany}

\author{Sudarsan Surendralal}
\affiliation[MPIE]
{Max-Planck-Institut f\"ur Eisenforschung GmbH, Max-Planck-Stra{\ss}e 1, 40237 D\"usseldorf, Germany}

\author{Mira Todorova}
\affiliation[MPIE]
{Max-Planck-Institut f\"ur Eisenforschung GmbH, Max-Planck-Stra{\ss}e 1, 40237 D\"usseldorf, Germany}

\author{Stefan Wippermann}
\email{stefan.wippermann@physik.uni-marburg.de}
\affiliation[MPIE]
{Max-Planck-Institut f\"ur Eisenforschung GmbH, Max-Planck-Stra{\ss}e 1, 40237 D\"usseldorf, Germany}
\alsoaffiliation[UMR]
{Philipps-Universit\"at Marburg, Renthof 5, 35032 Marburg, Germany}

\author{J\"org Neugebauer}
\affiliation[MPIE]
{Max-Planck-Institut f\"ur Eisenforschung GmbH, Max-Planck-Stra{\ss}e 1, 40237 D\"usseldorf, Germany}

\title[FSE]
{On the origin of univalent Mg$^+$ ions in solution and their role in anomalous anodic hydrogen evolution}
 
\begin{document}

\begin{abstract}
Aqueous metal corrosion is a major economic concern in modern society. A phenomenon that has puzzled generations of scientists in this field is the so-called anomalous hydrogen evolution: the violent dissolution of magnesium under electron-rich (anodic) conditions, accompanied by strong hydrogen evolution, and a key mechanism hampering Mg technology. Experimental studies have indicated the presence of univalent Mg$^+$ in solution, but these findings have been largely ignored because they defy our common chemical understanding and evaded direct experimental observation. Using recent advances in the \emph{ab initio} description of solid-liquid electrochemical interfaces under controlled potential conditions, we described the full reaction path of Mg atom dissolution from a kinked Mg surface under anodic conditions. Our study reveals the formation of a solvated [Mg$^{2+}$(OH)$^-$]$^+$ ion complex, challenging the conventional assumption of Mg$^{2+}$ ion. This insight provides an intuitive explanation for the postulated presence of (coulombically) univalent Mg$^+$ ions and the absence of protective oxide/hydroxide layers normally formed under anodic/oxidizing conditions. The discovery of this unexpected and unconventional reaction mechanism is crucial for identifying new strategies for corrosion prevention and can be transferred to other metals.
\end{abstract}


Controlling materials degradation in chemically harsh environments is an outstanding challenge for future sustainable technologies. Examples are electrochemical energy conversion and storage solutions \cite{ciamician,jaramillo,supercap1,mab2,transient1,transient2}, green metallurgy \cite{plasma,direct} and lightweight structural materials \cite{makar1993,esmaily2017fundamentals}. The need to understand the fundamental corrosion mechanisms in such environments is highlighted by a number of deceptively simple, yet poorly understood degradation reactions such as, e.g., the anomalous dissolution of metals under anodic conditions: their precise mechanistic details have remained elusive since their discovery more than 150 years ago \cite{beetz1866london}.

Magnesium is a prototypical example \cite{atrens2003}. Due to their light weight, high abundance and low environmental impact Mg alloys are attractive materials for mechanical engineering or batteries \cite{abbott2015magnesium}. However, with Mg being one of the most reactive metals a major technical weakness is corrosion in water. Many of the properties of magnesium in water are puzzling and not understood: For example, under anodic polarization, where Magnesium dissolves, it simultaneously shows extreme rates of hydrogen evolution (HE), which would normally be expected exclusively for cathodic potentials. According to the Butler-Volmer equation, HE should decrease exponentially when increasing the potential towards the anodic direction. In marked contrast, however, Magnesium and its alloys feature strongly enhanced HE with increasing anodic polarization. This effect is referred to in the literature as the 'negative difference effect' 
or 'anomalous hydrogen evolution reaction' \cite{beetz1866london,glicksman1959anodic,esmaily2017fundamentals}. Multiple models have been proposed \cite{birbilismechanisms2013,Liao2024} to explain the origin of the anomalous hydrogen evolution. The 'enhanced catalytic activity mechanism' \cite{Fajardo2016} suggested the enrichment of impurities more noble than Mg or the formation of local active sites to catalyze the HE on the anode. However, the specific nature of the catalyst that drives the anodic HE has remained unknown \cite{Fajardo2016,mercier2018}.

\begin{figure}[h]
  \centering
    \includegraphics[width=0.43\textwidth]{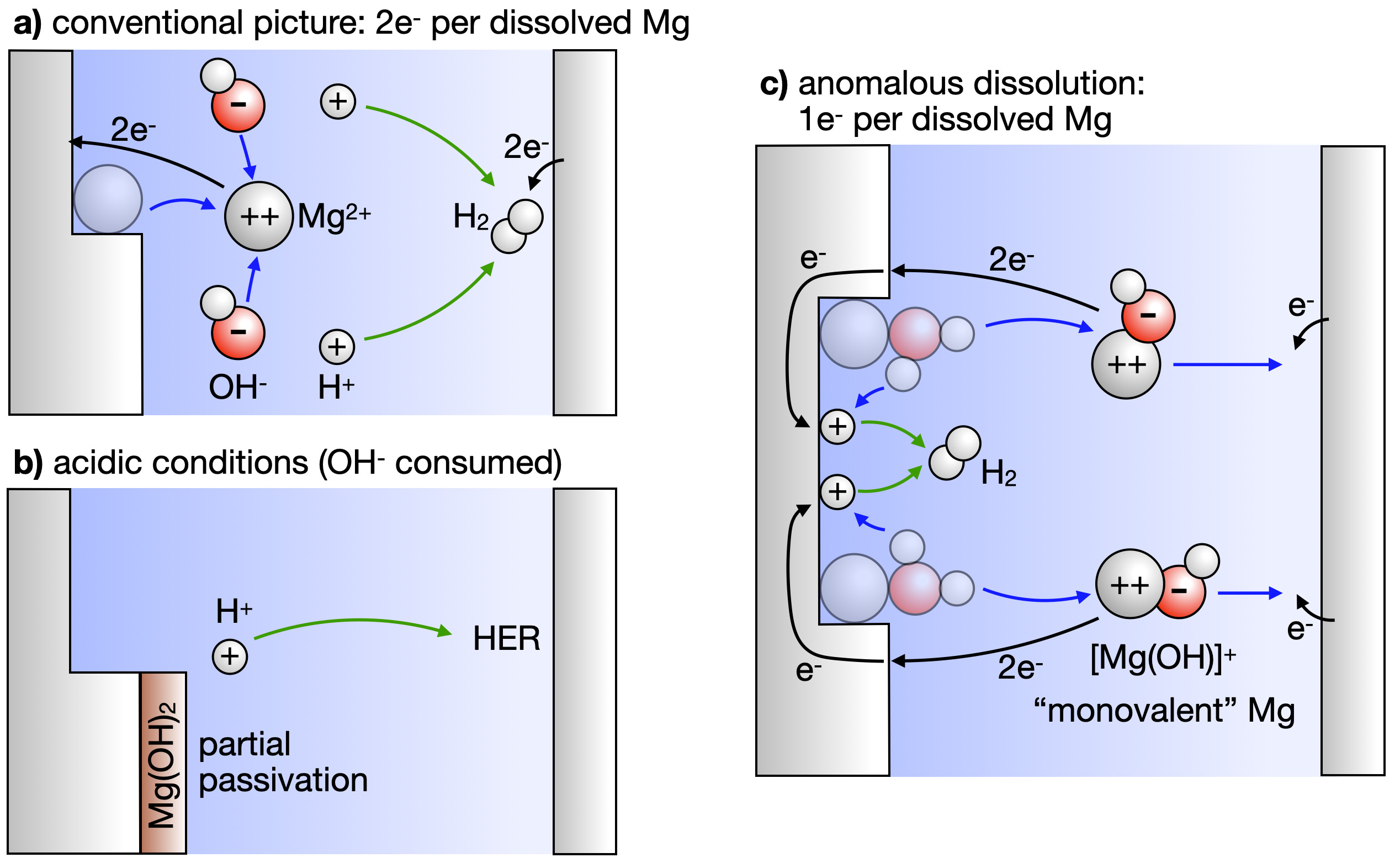}
\caption{\label{fig1} \textbf{a)} conventional picture of Mg corrosion: dissolution proceeds via formation of divalent Mg$^{2+}$ ions, resulting in the transfer of 2 e$^-$ per dissolved Mg ion. In solution, Mg$^{2+}$ reacts with OH$^-$ and precipitates. \textbf{b)} Under acidic conditions, OH$^-$ may be consumed, forming a hydroxide layer that partially passivates the Mg surface. \textbf{c)} In the present work, we reveal a low barrier pathway where Mg dissolves as an effectively monovalent [Mg$^{2+}$(OH)$^-$]$^+$ ion complex.}
\end{figure}

Even more puzzling, the amount of Mg dissolved was observed to be greater than coulometrically expected, assuming that Mg was oxidized to the dipositive Mg$^{2+}$ ion \cite{mgplus1}, cf. Fig. \ref{fig1}a. These findings were taken as evidence for the existence of a 'unipositive Mg$^+$ ion mechanism' \cite{mgplus2}. In a series of follow-up experiments, it was demonstrated that the unipositive Mg ions have a lifetime of several minutes in aqueous solutions and are able to reduce other species even macroscopic distances away from the oxidizing Mg anode \cite{mgplus1,mgplus3}. On the other hand, the unipositive Mg$^+$ mechanism has been challenged on the grounds that such an ion should be extremely short-lived. So far, there is only indirect evidence for the existence of Mg$^+$ \cite{esmaily2017fundamentals}. Atomic emission spectroelectrochemical experiments \cite{swiat2010,lebouil2014}, which clearly distinguish different oxidation states, found direct evidence only for divalent Mg$^{2+}$. Yet, this anomalous dissolution behaviour is not limited to Mg, but has been observed, e.g., for Fe, Cr and Zn as well \cite{drazic05}. Despite their fundamental importance, the existence and exact chemical nature of the postulated unipositive metal ions as well as the precise atomistic reaction mechanisms responsible for the anomalous dissolution, however, have remained elusive.

First principles techniques could be the method of choice to reveal the origin of these anomalous dissolution reactions. However, studies that explore the corrosion process taking into account the full complexity of the realistic surface-water interface are still lacking.

Via \emph{ab initio} thermopotentiostat molecular dynamics simulations, we demonstrate that the hypothesized unipositive metal ions are in fact ion complexes, consisting of a divalent metal ion and an OH$^-$ group. In the conventional picture, cf. Fig. \ref{fig1}a, Mg dissolves via the formation of divalent Mg$^{2+}$ ions, transferring 2e$^-$ to the metal surface per dissolved Mg ion. Under acidic conditions, cf. Fig. \ref{fig1}b, water adsorbs dissociatively, partially passivating the metal surface via formation of a Mg(OH)$_2$ surface hydroxide. In the present work, we identified an energetically and kinetically favourable reaction pathway where a surface metal atom becomes solvated in conjunction with an attached surface hydroxyl group (Fig. \ref{fig1}c). This pathway completely circumvents the passivating nature of the hydroxide film. In turn, this reaction process supports a continued dissociative water adsorption, explaining the anomalous hydrogen evolution at the anodically polarized surface. The effectively unipositive [Mg$^{2+}$(OH)$^-$]$^+$ ion complex is responsible for the observed amount of metal dissolved being larger than coulometrically expected. The [Mg$^{2+}$(OH)$^-$]$^+$ ion complex may subsequently decay into the divalent metal ion via reduction of another species. We expect it to be long-lived due to a strong Coulomb barrier separating the effectively unipositive ion complex from prospective reactants such as, e.g., H$^+$.

\begin{figure}[t]
  \centering
    \includegraphics[width=0.48\textwidth]{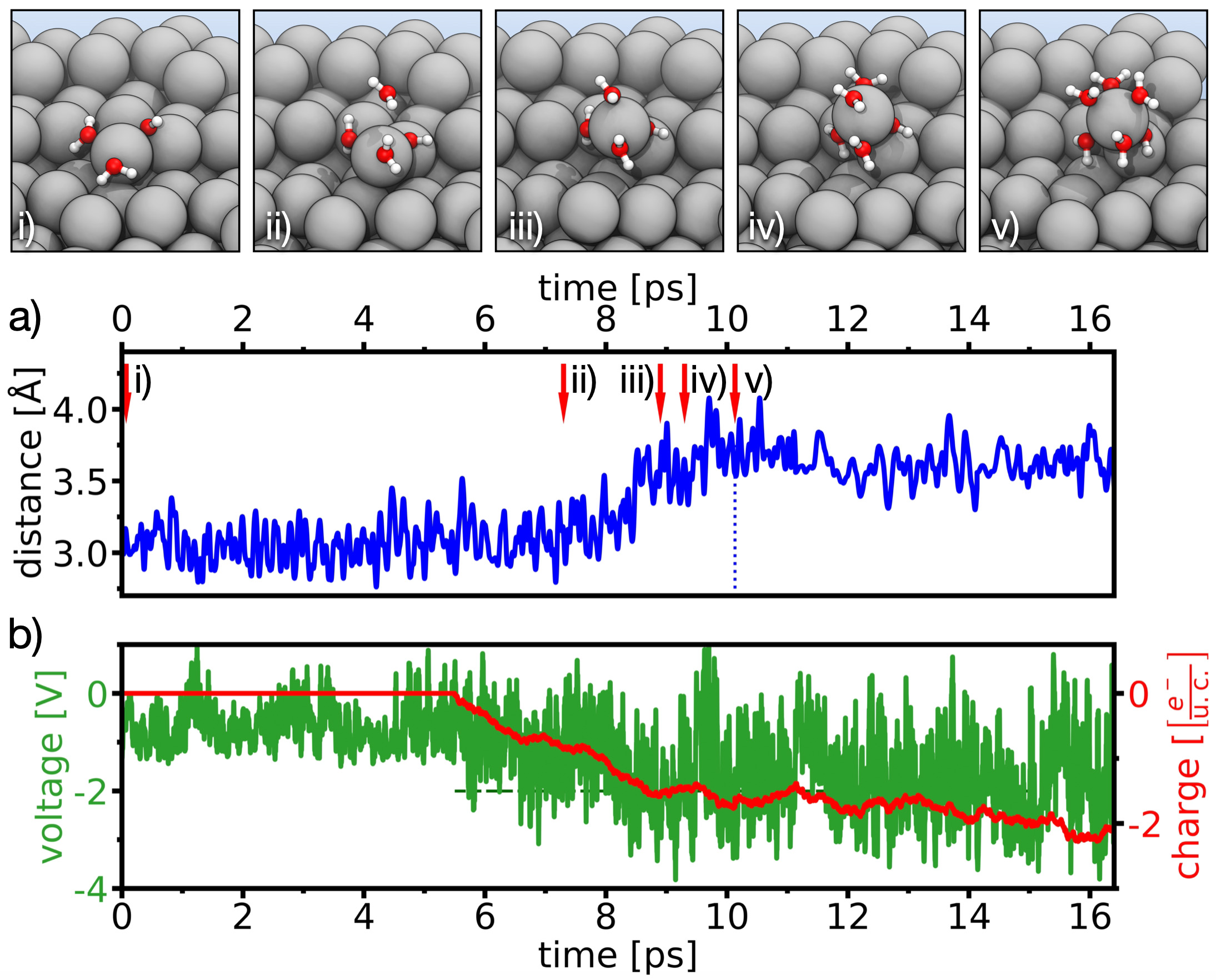}
\caption{\label{mgsolv} Solvation shell formation at kink-site Mg: \textbf{a)} Normal distance between the dissolving kink-site Mg-atom and its Mg bonding partner located directly underneath. The upper panel shows schematic representations of the structural evolution along the AIMD trajectory, at times marked by red arrows. For clarity, only water molecules belonging to the solvation shell of the Mg$^{2+}$ ion are shown explicitly. \textbf{b)} Local system potential (green curve) and counter charge (red curve) used to balance the surface charge on the working electrode. Potential control is turned on at time $t = 5.5$ ps. The green dashed line marks the targeted potential of \mbox{$\langle\Phi(t)\rangle = -2$ V.} The Mg$^{2+}$ ion remains attached to the surface via a hydroxyl bridge, see text.}
\end{figure}

To reveal the dissolution mechanism and the nature of the unipositive Mg$^+$ ion, we described the solvated Mg-surface by a supercell containing a 6-layer slab oriented in the (1 2 $\bar{3}$ 15) direction and 64 explicit water molecules. Dissolution is generally understood to proceed via kink atoms due to their weaker  bonds and the greater exposure of kink-sites to both adsorbing molecules and to the electric field. We therefore induced a miscut in the Mg-slab resulting in a surface with two kink-sites in the supercell. Already during equilibration under open-circuit conditions, a water molecule adsorbes dissociatively at one of the kink-sites according to:

\begin{equation}
Mg + H_2O_{(ad)} \rightleftharpoons [Mg^{2+}(OH)^-]^{+}_{(ad)} + \frac{H_{2_{(ad)}}}{2} + e^- \label{adsorb}
\end{equation}
Two further H$_2$O adsorbed subsequently at the same kink-site as intact molecules, leading to the configuration shown schematically in Fig. \ref{mgsolv}i. For clarity, only the participating water molecules are shown. Under open-circuit conditions, this configuration remained stable on the time-scale of our simulations.

\begin{figure}[t]
  \centering
    \includegraphics[width=0.48\textwidth]{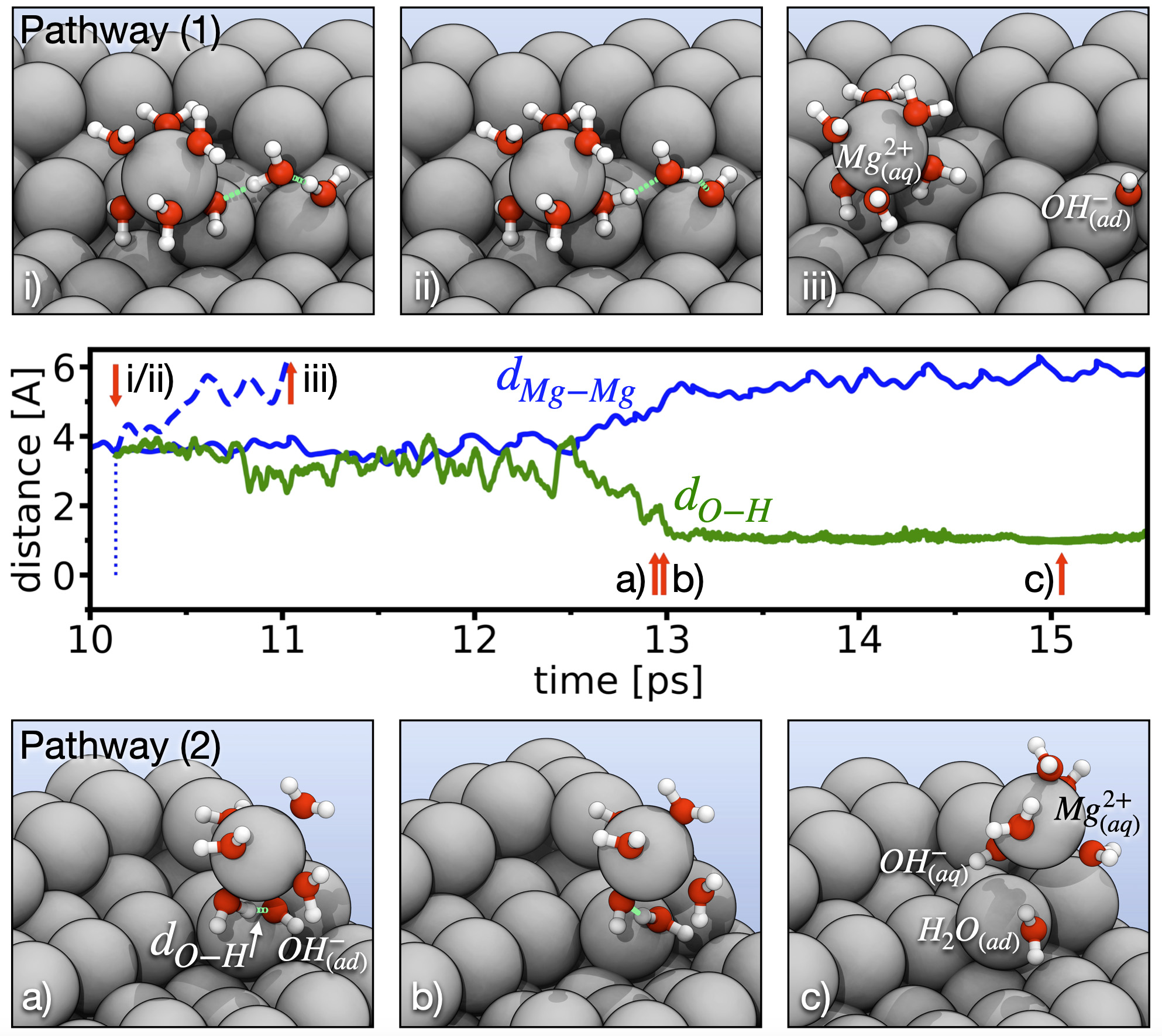}
\caption{\label{mgdiss} Two distinct pathways for Mg-dissolution: \textbf{(1)} A concerted double proton transfer from a surface adsorbed H$_2$O molecule to the hydroxyl bridge releases the Mg$^{2+}$ ion into solution. The corresponding normal distance between the dissolving kink-site Mg-atom and its Mg bonding partner underneath is indicated by the blue dashed line. The hydroxyl group remains on the surface. \textbf{(2)} Alternatively, an intra solvation shell single proton transfer to the hydroxyl bridge equally detaches the Mg$^{2+}$ ion from the surface (solid blue line). The green line denotes the distance betwen the transferring proton and the hydroxyl bridge. The hydroxyl remains attached to the Mg$^{2+}$ ion, resulting in an effectively +1 charged [Mg$^{2+}$OH$^{-}$]$^+$ ion complex.}
\end{figure}

In order to drive a dissolution reaction, we subsequently polarized the Mg slab anodically. The electrode charge is controlled by our recently introduced thermopotentiostat \cite{deissenbeck, deissenbeck2}. Within 1.6 ps after switching on the thermopotentiostat with a target potential of $\langle\Phi\rangle = -2$ V, a fourth water molecule approached the kink atom (cf. Fig. \ref{mgsolv}ii) and adsorbed (Fig. \ref{mgsolv}iii), starting to form a solvation shell. The kink-atom is then increasingly being lifted out of the surface. Fig. \ref{mgsolv}a shows the distance parallel to the surface normal between the kink-atom and its Mg bonding partner underneath. A maximum extension of 3.6 \AA\ is reached after completing the solvation shell (Fig. \ref{mgsolv}iv-v). Although the surface is charged with 2 additional electrons (cf. Fig. \ref{mgsolv}b), indicating that the kink atom is now fully ionized, the solvated Mg$^{2+}$ ion remains firmly bound to the surface: in conjunction with the hydroxyl group created in the reaction (\ref{adsorb}), the solvated ion forms an [Mg$^{2+}$OH$^-$]$^+$ complex, where the hydroxyl group connects the kink atom to its nearest Mg neighbour (cf. Fig. \ref{mgsolv}v). We speculate that this process is common to anodically polarized metal surfaces. A 'place-exchange mechanism', where a surface-adsorbed oxygen and a metal atom underneath exchange their position, was first proposed by Lanyon \emph{et. al.} \cite{lanyon1955}, later observed by Vetter \emph{et al.}\cite{vetter1972} for Pt surfaces and reexamined more recently by Rost \emph{et al.} \cite{rost2023}.

This hydroxyl bridge bond is highly stable. Removing any water molecules except the ones constituting the solvation shell and separating the [Mg$^{2+}$OH$^-$]$^+$ ion complex from the surface in a vacuum calculation, we estimated the binding energy to be $\sim$ 2 eV. Such a large binding energy is inconsistent with the experimentally observed high dissolution rates \cite{beetz1866london,glicksman1959anodic,esmaily2017fundamentals}. In order for the dissolution to proceed, we therefore expect the breaking of the hydroxyl bridge bond to be catalyzed by its surrounding environment. Indeed, our simulations showed a possible candidate: a concerted double proton transfer from a neighbouring adsorbed water molecule via a solvated H$_2$O molecule that is hydrogen bridge bonded to the hydroxyl group (cf. Fig. \ref{mgdiss}i) relocates the hydroxyl group laterally to a neighbouring site. Thereby, the [Mg$^{2+}$OH$^-$]$^+$ ion complex is oxidized to Mg$^{2+}$ and left with a complete solvation shell consisting of 6 H$_2$O molecules (Fig. \ref{mgdiss}ii): 
\begin{equation}
[Mg^{2+}(OH)^-]^+_{(ad)} + H_2O_{(ad)} \rightleftharpoons Mg^{2+}_{(aq)} + OH^-_{(ad)} + H_2O_{(aq)} \label{diss1}
\end{equation}
As a result, the Mg$^{2+}$ ion quickly moved into the liquid water region (dashed blue line in Fig. \ref{mgsolv}), leaving the hydroxyl group behind on the surface.

This double proton transfer proceeding on the surface is, however, not the only conceivable reaction to catalyze the dissolution. In order to search for alternative processes, we moved the Mg$^{2+}$ ion with a constant velocity of  $v$ = 2/3 \AA/ps parallel to the surface normal into the solution. In response, one of the H$_2$O molecules forming the solvation shell turned one of its OH-bonds towards the hydroxyl group. In Fig. \ref{mgdiss}, we show the distance between the hydrogen in the corresponding OH-bond and the oxygen atom of the hydroxyl group (green solid curve). In the time frame from 10.8 ps to 12.5 ps, multiple transfer attempts are visible, until at $\sim$ 13 ps an intra solvation shell single proton transfer occurs to the hydroxyl group (Fig. \ref{mgdiss}a/b). Thereby, the [Mg$^{2+}$OH$^-$]$^+$ ion complex as a whole becomes fully solvated and moves into the liquid water region (solid blue curve, Fig. \ref{mgdiss}c):
\begin{equation}
[Mg^{2+}(OH)^-]^+_{(ad)} + H_2O_{(aq)} \rightleftharpoons [Mg^{2+}(OH)^-]^+_{(aq)} + H_2O_{(ad)} \label{diss2}
\end{equation}
We emphasize, that the outcome of these two competing processes is fundamentally different. For reaction (\ref{diss1}), the hydroxyl remains on the surface. Therefore, the surface will be quickly hydroxylated and become electrochemically passive. No more dissociative H$_2$O adsorption is possible, preventing any further anomalous hydrogen evolution. Reaction (\ref{diss2}), on the other hand, removes the hydroxyl group from the surface into the solution, leaving the next kink site exposed to further dissociative H$_2$O adsorption. It is therefore only reaction (\ref{diss2}) that is associated with ongoing anomalous hydrogen evolution. Consistent with the experimental observation that only the unipositive Mg ion is associated with the anomalous anodic hydrogen evolution \cite{mgplus1,mgplus2}, the [Mg$^{2+}$OH$^-$]$^+$ ion complex created in reaction (\ref{diss2}) is effectively charged +1. We therefore propose that the elusive unipositive Mg ion is, in fact, an [Mg$^{2+}$OH$^-$]$^+$ ion complex. This interpretation is supported by the fact that the reaction $Mg^{2+}(OH)^- \rightleftharpoons Mg^{2+} + OH^-$ is known to have a $p_{K_b}$ value of 2.56 \cite{CRC}. Hence, for $pH > 11.44$ the [Mg$^{2+}$OH$^-$]$^+$ ion complex becomes the dominant species. Due to the hydroxylation of the surface, we speculate that the local $pH$ becomes sufficiently large, so that the ion complex may even be thermodynamically stabilized.

Moreover, Refs. \cite{mgplus1,mgplus3} pointed out that the unipositive Mg ion remains stable for several minutes in aqueous solutions and is able to reduce other species even macroscopic distances away from the Mg anode. Since the [Mg$^{2+}$OH$^-$]$^+$ ion complex is positively charged and requires, e.g., an H$^+$ ion to oxidize to Mg$^{2+}$, we expect the complex to be able to reduce other species. In addition, due to the Coulomb barrier between [Mg$^{2+}$OH$^-$]$^+$ and H$^+$, which both are positively charged, the ion complex will be rather long-lived. An alternative reaction to obtain Mg$^{2+}$ is the dissociation of the [Mg$^{2+}$OH$^-$]$^+$ ion complex into its constituents Mg$^{2+}$ and OH$^-$. Analogous to the reduction via other species, this reaction is kinetically hindered by the large Coulomb attraction between the positive Mg$^{2+}$ and the negative OH$^-$.

\begin{figure}[t]
  \centering
    \includegraphics[width=0.48\textwidth]{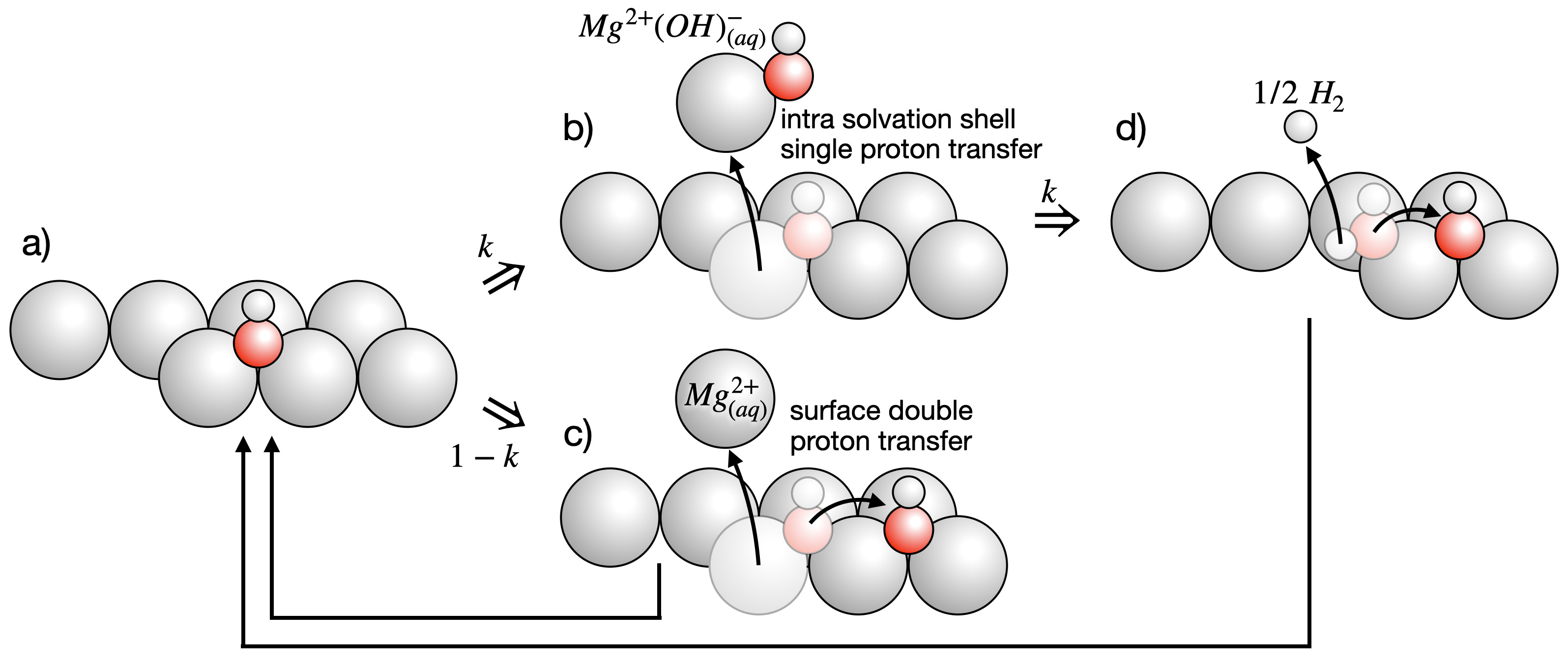}
\caption{\label{schem} Reaction model for Mg dissolution and anomalous hydrogen evolution via unipositive Mg, see text.}
\end{figure}

The reaction steps observed in our thermopotentiostat \emph{AIMD} simulations imply the following model for Mg dissolution and the anomalous HER via unipositive Mg: starting from the hydroxylated surface (Fig. \ref{schem}a), two distinct reaction pathways are available. On the one hand, with a rate of $1-k$ the hydroxylated kink atom is first ionized as
\begin{equation}
(1-k) \cdot \left[ Mg + OH^- \rightleftharpoons [Mg^{2+}(OH)^-]^+ + 2 e^- \right] \label{stepa}
\end{equation}
and subsequently solvated according to (Fig. \ref{schem}c):
%
%
\begin{equation}
(1-k) \cdot \left[ [Mg^{2+}(OH)^-]^+ \rightleftharpoons Mg^{2+} + OH^- \right] \label{stepc}
\end{equation}

After that, the process can start over at the next exposed kink atom (return to Fig. \ref{schem}a). We note that summing Eqs. \ref{stepa} and \ref{stepc} results in
\begin{equation}
(1-k) \cdot \left[ Mg \rightleftharpoons Mg^{2+} + 2e^- \right]. \label{sum1}
\end{equation}

On the other hand, with a rate of $k$ the $[Mg^{2+}(OH)^-]^+$ ion complex is solvated as a whole (Fig. \ref{schem}b). Since the hydroxyl group becomes detached from the surface, the surface is thereby left exposed to another dissociative adsorption event (Fig. \ref{schem}d):
\begin{equation}
k \cdot \left[ Mg + H_2O \rightleftharpoons [Mg^{2+}(OH)^-]^+ + \frac{H_2}{2} + e^- \right] \label{stepd}
\end{equation}
This is the step that triggers the anomalous anodic hydrogen evolution reaction (HER), explaining why only the effectively unipositive Mg ion complex contributes to the anodic HER.

Eventually, the $[Mg^{2+}(OH)^-]^+$ complexes in the anolyte reduce other species - such as, e.g., H$^+$ - and oxidize in the process according to:
\begin{equation}
k \cdot \left[ [Mg^{2+}(OH)^-]^+ + H^+ \rightleftharpoons Mg^{2+} + H_2O \right] \label{step_ox}
\end{equation}
Summing Eqs. \ref{stepd} and \ref{step_ox} yields:
\begin{equation}
k \cdot \left[ Mg + H^+ \rightleftharpoons Mg^{2+} + \frac{H_2}{2} + e^- \right] \label{sum2}
\end{equation}
We now see that Eqs. \ref{sum1} and \ref{sum2} are balanced by the cathodic half-reaction
\begin{equation}
(2 - k) \cdot \left[H^+ + e^- \rightleftharpoons \frac{H_2}{2} \right], \label{step_cath}
\end{equation}
so that the sum of Eqs. \ref{sum1}, \ref{sum2} and \ref{step_cath} yields the well known total balance equation
\begin{equation}
Mg + 2 H^+ \rightleftharpoons Mg^{2+} + H_2.
\end{equation}
We emphasize, that the key steps \ref{stepa}, \ref{stepc} and \ref{stepd} are directly observable in our AIMD simulations. Only steps \ref{step_ox} and \ref{step_cath} have been inferred.

In summary, by using \emph{ab initio} molecular dynamics simulations of aqueous magnesium interfaces under potential control, taking into account the full complexity of the realistic metal-water interface, we have discovered a novel and completely unexpected reaction pathway. The identified dissolution product - $[Mg^{2+}OH^-]^+$ - naturally explains one of the most studied and debated corrosion mechanisms - the anomalous anodic hydrogen evolution, which has puzzled scientists since it was first reported more than 150 years ago. 
Our results clearly show that water is not just a spectator, but an active reactant. Under anodic conditions, water dissociatively adsorbs to form a surface hydroxide. Subsequently, the interfacial water provides a low-barrier pathway for proton transfer reactions that allow the surface hydroxide to dissolve via the formation of [Mg$^{2+}$(OH)$^-$]$^+$ ion complexes. 
This pathway bypasses the usual passivation effect of surface films and explains the unusually high anodic corrosion rates and the chemical nature of the hypothesized solvated Mg$^+$ ions. 
The discovery of such an unexpected reaction pathway also demonstrates the level and potential that ab initio molecular dynamics simulations have reached thanks to recent methodological advances in the description of electrochemical interfaces.  

The authors acknowledge the Deutsche Forschungsgemeinschaft (DFG, German Research Foundation) for funding through Project No. 409476157 (SFB1394) and support under Germany's Excellence Strategy - EXC 2033-390677874–RESOLV. 

\bibliography{main.bib}

\providecommand{\latin}[1]{#1}
\makeatletter
\providecommand{\doi}
  {\begingroup\let\do\@makeother\dospecials
  \catcode`\{=1 \catcode`\}=2 \doi@aux}
\providecommand{\doi@aux}[1]{\endgroup\texttt{#1}}
\makeatother
\providecommand*\mcitethebibliography{\thebibliography}
\csname @ifundefined\endcsname{endmcitethebibliography}
  {\let\endmcitethebibliography\endthebibliography}{}
\begin{mcitethebibliography}{31}
\providecommand*\natexlab[1]{#1}
\providecommand*\mciteSetBstSublistMode[1]{}
\providecommand*\mciteSetBstMaxWidthForm[2]{}
\providecommand*\mciteBstWouldAddEndPuncttrue
  {\def\EndOfBibitem{\unskip.}}
\providecommand*\mciteBstWouldAddEndPunctfalse
  {\let\EndOfBibitem\relax}
\providecommand*\mciteSetBstMidEndSepPunct[3]{}
\providecommand*\mciteSetBstSublistLabelBeginEnd[3]{}
\providecommand*\EndOfBibitem{}
\mciteSetBstSublistMode{f}
\mciteSetBstMaxWidthForm{subitem}{(\alph{mcitesubitemcount})}
\mciteSetBstSublistLabelBeginEnd
  {\mcitemaxwidthsubitemform\space}
  {\relax}
  {\relax}

\bibitem[Ciamician(1912)]{ciamician}
Ciamician,~G. The photochemistry of the future. \emph{Science} \textbf{1912},
  \emph{36}, 385--394\relax
\mciteBstWouldAddEndPuncttrue
\mciteSetBstMidEndSepPunct{\mcitedefaultmidpunct}
{\mcitedefaultendpunct}{\mcitedefaultseppunct}\relax
\EndOfBibitem
\bibitem[Seh \latin{et~al.}(2017)Seh, Kibsgaard, Dickens, Chorkendorff,
  Norskov, and Jaramillo]{jaramillo}
Seh,~Z.~W.; Kibsgaard,~J.; Dickens,~C.~F.; Chorkendorff,~I.; Norskov,~J.;
  Jaramillo,~T.~F. Combining theory and experiment in electrocatalysis:
  Insights into materials design. \emph{Science} \textbf{2017}, \emph{355},
  146\relax
\mciteBstWouldAddEndPuncttrue
\mciteSetBstMidEndSepPunct{\mcitedefaultmidpunct}
{\mcitedefaultendpunct}{\mcitedefaultseppunct}\relax
\EndOfBibitem
\bibitem[Wang \latin{et~al.}(2016)Wang, Song, and Xia]{supercap1}
Wang,~Y.; Song,~Y.; Xia,~Y. Electrochemical capacitors: mechanism, materials,
  systems, characterization and applications. \emph{CHEMICAL SOCIETY REVIEWS}
  \textbf{2016}, \emph{45}, 5925--5950\relax
\mciteBstWouldAddEndPuncttrue
\mciteSetBstMidEndSepPunct{\mcitedefaultmidpunct}
{\mcitedefaultendpunct}{\mcitedefaultseppunct}\relax
\EndOfBibitem
\bibitem[Suntivich \latin{et~al.}(2011)Suntivich, Gasteiger, Yabuuchi,
  Nakanishi, Goodenough, and Shao-Horn]{mab2}
Suntivich,~J.; Gasteiger,~H.~A.; Yabuuchi,~N.; Nakanishi,~H.;
  Goodenough,~J.~B.; Shao-Horn,~Y. Design principles for oxygen-reduction
  activity on perovskite oxide catalysts for fuel cells and metal-air
  batteries. \emph{NATURE CHEMISTRY} \textbf{2011}, \emph{3}, 546--550\relax
\mciteBstWouldAddEndPuncttrue
\mciteSetBstMidEndSepPunct{\mcitedefaultmidpunct}
{\mcitedefaultendpunct}{\mcitedefaultseppunct}\relax
\EndOfBibitem
\bibitem[Hwang \latin{et~al.}(2012)Hwang, Tao, Kim, Cheng, Song, Rill,
  Brenckle, Panilaitis, Won, Kim, Song, Yu, Ameen, Li, Su, Yang, Kaplan, Zakin,
  Slepian, Huang, Omenetto, and Rogers]{transient1}
Hwang,~S.-W. \latin{et~al.}  A Physically Transient Form of Silicon
  Electronics. \emph{SCIENCE} \textbf{2012}, \emph{337}, 1640--1644\relax
\mciteBstWouldAddEndPuncttrue
\mciteSetBstMidEndSepPunct{\mcitedefaultmidpunct}
{\mcitedefaultendpunct}{\mcitedefaultseppunct}\relax
\EndOfBibitem
\bibitem[Yu \latin{et~al.}(2016)Yu, Kuzum, Hwang, Kim, Juul, Kim, Won, Chiang,
  Trumpis, Richardson, Cheng, Fang, Thompson, Bink, Talos, Seo, Lee, Kang, Kim,
  Lee, Huang, Jensen, Dichter, Lucas, Viventi, Litt, and Rogers]{transient2}
Yu,~K.~J. \latin{et~al.}  Bioresorbable silicon electronics for transient
  spatiotemporal mapping of electrical activity from the cerebral cortex.
  \emph{NATURE MATERIALS} \textbf{2016}, \emph{15}, 782+\relax
\mciteBstWouldAddEndPuncttrue
\mciteSetBstMidEndSepPunct{\mcitedefaultmidpunct}
{\mcitedefaultendpunct}{\mcitedefaultseppunct}\relax
\EndOfBibitem
\bibitem[Sabat \latin{et~al.}(2014)Sabat, Rajput, Paramguru, Bhoi, and
  Mishra]{plasma}
Sabat,~K.~C.; Rajput,~P.; Paramguru,~R.~K.; Bhoi,~B.; Mishra,~B.~K. Reduction
  of Oxide Minerals by Hydrogen Plasma: An Overview. \emph{PLASMA CHEMISTRY AND
  PLASMA PROCESSING} \textbf{2014}, \emph{34}, 1--23\relax
\mciteBstWouldAddEndPuncttrue
\mciteSetBstMidEndSepPunct{\mcitedefaultmidpunct}
{\mcitedefaultendpunct}{\mcitedefaultseppunct}\relax
\EndOfBibitem
\bibitem[Cavaliere(2022)]{direct}
Cavaliere,~P. \emph{Hydrogen Assisted Direct Reduction of Iron Oxides};
  Springer Cham, 2022\relax
\mciteBstWouldAddEndPuncttrue
\mciteSetBstMidEndSepPunct{\mcitedefaultmidpunct}
{\mcitedefaultendpunct}{\mcitedefaultseppunct}\relax
\EndOfBibitem
\bibitem[Makar and Kruger(1993)Makar, and Kruger]{makar1993}
Makar,~G.~L.; Kruger,~J. Corrosion of magnesium. \emph{International Materials
  Reviews} \textbf{1993}, \emph{38}, 138--153\relax
\mciteBstWouldAddEndPuncttrue
\mciteSetBstMidEndSepPunct{\mcitedefaultmidpunct}
{\mcitedefaultendpunct}{\mcitedefaultseppunct}\relax
\EndOfBibitem
\bibitem[Esmaily \latin{et~al.}(2017)Esmaily, Svensson, Fajardo, Birbilis,
  Frankel, Virtanen, Arrabal, Thomas, and Johansson]{esmaily2017fundamentals}
Esmaily,~M.; Svensson,~J.; Fajardo,~S.; Birbilis,~N.; Frankel,~G.;
  Virtanen,~S.; Arrabal,~R.; Thomas,~S.; Johansson,~L. Fundamentals and
  advances in magnesium alloy corrosion. \emph{Progress in Materials Science}
  \textbf{2017}, \emph{89}, 92--193\relax
\mciteBstWouldAddEndPuncttrue
\mciteSetBstMidEndSepPunct{\mcitedefaultmidpunct}
{\mcitedefaultendpunct}{\mcitedefaultseppunct}\relax
\EndOfBibitem
\bibitem[Beetz(1866)]{beetz1866london}
Beetz,~W. London, Edinburgh, Dublin Philos. Mag. \emph{J. Sci} \textbf{1866},
  \emph{32}, 269\relax
\mciteBstWouldAddEndPuncttrue
\mciteSetBstMidEndSepPunct{\mcitedefaultmidpunct}
{\mcitedefaultendpunct}{\mcitedefaultseppunct}\relax
\EndOfBibitem
\bibitem[Song and Atrens(2003)Song, and Atrens]{atrens2003}
Song,~G.; Atrens,~A. Understanding Magnesium Corrosion—A Framework for
  Improved Alloy Performance. \emph{Advanced Engineering Materials}
  \textbf{2003}, \emph{5}, 837--858\relax
\mciteBstWouldAddEndPuncttrue
\mciteSetBstMidEndSepPunct{\mcitedefaultmidpunct}
{\mcitedefaultendpunct}{\mcitedefaultseppunct}\relax
\EndOfBibitem
\bibitem[Abbott(2015)]{abbott2015magnesium}
Abbott,~T.~B. Magnesium: industrial and research developments over the last 15
  years. \emph{Corrosion} \textbf{2015}, \emph{71}, 120--127\relax
\mciteBstWouldAddEndPuncttrue
\mciteSetBstMidEndSepPunct{\mcitedefaultmidpunct}
{\mcitedefaultendpunct}{\mcitedefaultseppunct}\relax
\EndOfBibitem
\bibitem[Glicksman(1959)]{glicksman1959anodic}
Glicksman,~R. Anodic dissolution of Magnesium alloys in aqueous salt solutions.
  \emph{Journal of the Electrochemical Society} \textbf{1959}, \emph{106},
  83\relax
\mciteBstWouldAddEndPuncttrue
\mciteSetBstMidEndSepPunct{\mcitedefaultmidpunct}
{\mcitedefaultendpunct}{\mcitedefaultseppunct}\relax
\EndOfBibitem
\bibitem[Frankel \latin{et~al.}(2013)Frankel, Samaniego, and
  Birbilis]{birbilismechanisms2013}
Frankel,~G.; Samaniego,~A.; Birbilis,~N. Evolution of hydrogen at dissolving
  magnesium surfaces. \emph{Corrosion Science} \textbf{2013}, \emph{70},
  104--111\relax
\mciteBstWouldAddEndPuncttrue
\mciteSetBstMidEndSepPunct{\mcitedefaultmidpunct}
{\mcitedefaultendpunct}{\mcitedefaultseppunct}\relax
\EndOfBibitem
\bibitem[Liao \latin{et~al.}(2024)Liao, Sun, Huang, Xiao, Ying, and
  Cao]{Liao2024}
Liao,~X.; Sun,~C.; Huang,~J.; Xiao,~B.; Ying,~T.; Cao,~F. The positive
  difference effect and anomalous hydrogen evolution on the anodically
  polarized Mg surface in alkaline NH4+ containing solutions. \emph{Journal of
  Materials Science} \textbf{2024}, \emph{59}, 8535--8555\relax
\mciteBstWouldAddEndPuncttrue
\mciteSetBstMidEndSepPunct{\mcitedefaultmidpunct}
{\mcitedefaultendpunct}{\mcitedefaultseppunct}\relax
\EndOfBibitem
\bibitem[Fajardo \latin{et~al.}(2016)Fajardo, Glover, Williams, and
  Frankel]{Fajardo2016}
Fajardo,~S.; Glover,~C.; Williams,~G.; Frankel,~G. The Source of Anodic
  Hydrogen Evolution on Ultra High Purity Magnesium. \emph{Electrochimica Acta}
  \textbf{2016}, \emph{212}, 510--521\relax
\mciteBstWouldAddEndPuncttrue
\mciteSetBstMidEndSepPunct{\mcitedefaultmidpunct}
{\mcitedefaultendpunct}{\mcitedefaultseppunct}\relax
\EndOfBibitem
\bibitem[Mercier \latin{et~al.}(2018)Mercier, Swiatkowska, Zanna, Seyeux, and
  Marcus]{mercier2018}
Mercier,~D.; Swiatkowska,~J.; Zanna,~S.; Seyeux,~A.; Marcus,~P. Role of
  Segregated Iron at Grain Boundaries on Mg Corrosion. \emph{Journal of the
  Electrochemical Society} \textbf{2018}, \emph{165}, C42\relax
\mciteBstWouldAddEndPuncttrue
\mciteSetBstMidEndSepPunct{\mcitedefaultmidpunct}
{\mcitedefaultendpunct}{\mcitedefaultseppunct}\relax
\EndOfBibitem
\bibitem[Petty \latin{et~al.}(1954)Petty, Davidson, and Kleinberg]{mgplus1}
Petty,~R.~L.; Davidson,~A.~W.; Kleinberg,~J. The Anodic Oxidation of Magnesium
  Metal: Evidence for the Existence of Unipositive Magnesium1,2. \emph{Journal
  of the American Chemical Society} \textbf{1954}, \emph{76}, 363--366\relax
\mciteBstWouldAddEndPuncttrue
\mciteSetBstMidEndSepPunct{\mcitedefaultmidpunct}
{\mcitedefaultendpunct}{\mcitedefaultseppunct}\relax
\EndOfBibitem
\bibitem[Atrens and Dietzel(2007)Atrens, and Dietzel]{mgplus2}
Atrens,~A.; Dietzel,~W. The Negative Difference Effect and Unipositive Mg+.
  \emph{Advanced Engineering Materials} \textbf{2007}, \emph{9}, 292--297\relax
\mciteBstWouldAddEndPuncttrue
\mciteSetBstMidEndSepPunct{\mcitedefaultmidpunct}
{\mcitedefaultendpunct}{\mcitedefaultseppunct}\relax
\EndOfBibitem
\bibitem[Rausch \latin{et~al.}(1957)Rausch, McEwen, and Kleinberg]{mgplus3}
Rausch,~M.~D.; McEwen,~W.~E.; Kleinberg,~J. Reductions Involving Unipositive
  Magnesium. \emph{Chemical Reviews} \textbf{1957}, \emph{57}, 417--437\relax
\mciteBstWouldAddEndPuncttrue
\mciteSetBstMidEndSepPunct{\mcitedefaultmidpunct}
{\mcitedefaultendpunct}{\mcitedefaultseppunct}\relax
\EndOfBibitem
\bibitem[Światowska \latin{et~al.}(2010)Światowska, Volovitch, and
  Ogle]{swiat2010}
Światowska,~J.; Volovitch,~P.; Ogle,~K. The anodic dissolution of Mg in NaCl
  and Na2SO4 electrolytes by atomic emission spectroelectrochemistry.
  \emph{Corrosion Science} \textbf{2010}, \emph{52}, 2372--2378\relax
\mciteBstWouldAddEndPuncttrue
\mciteSetBstMidEndSepPunct{\mcitedefaultmidpunct}
{\mcitedefaultendpunct}{\mcitedefaultseppunct}\relax
\EndOfBibitem
\bibitem[Lebouil \latin{et~al.}(2014)Lebouil, Gharbi, Volovitch, and
  Ogle]{lebouil2014}
Lebouil,~S.; Gharbi,~O.; Volovitch,~P.; Ogle,~K. {Mg Dissolution in Phosphate
  and Chloride Electrolytes: Insight into the Mechanism of the Negative
  Difference Effect}. \emph{Corrosion} \textbf{2014}, \emph{71}, 234--241\relax
\mciteBstWouldAddEndPuncttrue
\mciteSetBstMidEndSepPunct{\mcitedefaultmidpunct}
{\mcitedefaultendpunct}{\mcitedefaultseppunct}\relax
\EndOfBibitem
\bibitem[Drazic and Popic(2005)Drazic, and Popic]{drazic05}
Drazic,~D.~M.; Popic,~J.~P. Anomalous dissolution of metals and chemical
  corrosion. \emph{Journal of the Serbian Chemical Society} \textbf{2005},
  \emph{70}, 489--511\relax
\mciteBstWouldAddEndPuncttrue
\mciteSetBstMidEndSepPunct{\mcitedefaultmidpunct}
{\mcitedefaultendpunct}{\mcitedefaultseppunct}\relax
\EndOfBibitem
\bibitem[Deissenbeck \latin{et~al.}(2021)Deissenbeck, Freysoldt, Todorova,
  Neugebauer, and Wippermann]{deissenbeck}
Deissenbeck,~F.; Freysoldt,~C.; Todorova,~M.; Neugebauer,~J.; Wippermann,~S.
  Dielectric Properties of Nanoconfined Water: A Canonical Thermopotentiostat
  Approach. \emph{Phys. Rev. Lett.} \textbf{2021}, \emph{126}, 136803\relax
\mciteBstWouldAddEndPuncttrue
\mciteSetBstMidEndSepPunct{\mcitedefaultmidpunct}
{\mcitedefaultendpunct}{\mcitedefaultseppunct}\relax
\EndOfBibitem
\bibitem[Deißenbeck and Wippermann(2023)Deißenbeck, and
  Wippermann]{deissenbeck2}
Deißenbeck,~F.; Wippermann,~S. Dielectric Properties of Nanoconfined Water
  from Ab Initio Thermopotentiostat Molecular Dynamics. \emph{Journal of
  Chemical Theory and Computation} \textbf{2023}, \emph{19}, 1035--1043, PMID:
  36705611\relax
\mciteBstWouldAddEndPuncttrue
\mciteSetBstMidEndSepPunct{\mcitedefaultmidpunct}
{\mcitedefaultendpunct}{\mcitedefaultseppunct}\relax
\EndOfBibitem
\bibitem[Lanyon \latin{et~al.}(1955)Lanyon, Trapnell, and
  Hinshelwood]{lanyon1955}
Lanyon,~M. A.~H.; Trapnell,~B. M.~W.; Hinshelwood,~C.~N. The interaction of
  oxygen with clean metal surfaces. \emph{Proceedings of the Royal Society of
  London. Series A. Mathematical and Physical Sciences} \textbf{1955},
  \emph{227}, 387--399\relax
\mciteBstWouldAddEndPuncttrue
\mciteSetBstMidEndSepPunct{\mcitedefaultmidpunct}
{\mcitedefaultendpunct}{\mcitedefaultseppunct}\relax
\EndOfBibitem
\bibitem[Vetter and Schultze(1972)Vetter, and Schultze]{vetter1972}
Vetter,~K.~J.; Schultze,~J.~W. The kinetics of the electrochemical formation
  and reduction of monomolecular oxide layers on platinum in 0.5 M H2SO4: Part
  I. Potentiostatic pulse measurements. \emph{Journal of Electroanalytical
  Chemistry and Interfacial Electrochemistry} \textbf{1972}, \emph{34},
  131--139\relax
\mciteBstWouldAddEndPuncttrue
\mciteSetBstMidEndSepPunct{\mcitedefaultmidpunct}
{\mcitedefaultendpunct}{\mcitedefaultseppunct}\relax
\EndOfBibitem
\bibitem[Rost \latin{et~al.}(2023)Rost, Jacobse, and Koper]{rost2023}
Rost,~M.~J.; Jacobse,~L.; Koper,~M. T.~M. Non-Random Island Nucleation in the
  Electrochemical Roughening on Pt(111). \emph{Angewandte Chemie International
  Edition} \textbf{2023}, \emph{62}, e202216376\relax
\mciteBstWouldAddEndPuncttrue
\mciteSetBstMidEndSepPunct{\mcitedefaultmidpunct}
{\mcitedefaultendpunct}{\mcitedefaultseppunct}\relax
\EndOfBibitem
\bibitem[Haynes \latin{et~al.}(2017)Haynes, Lide, and Bruni]{CRC}
Haynes,~W.~M.; Lide,~D.~R.; Bruni,~T.~J. \emph{CRC Handbook of Chemistry and
  Physics}; Taylor and Francis, 2017\relax
\mciteBstWouldAddEndPuncttrue
\mciteSetBstMidEndSepPunct{\mcitedefaultmidpunct}
{\mcitedefaultendpunct}{\mcitedefaultseppunct}\relax
\EndOfBibitem
\end{mcitethebibliography}

\end{document}